# Absence of Spin Frustration in the Kagomé Layers of $Cu^{2+}$ Ions in Volborthite $Cu_3V_2O_7(OH)_2·2H_2O$ and Observation of the Suppression and Re-entrance of Specific Heat Anomalies in Volborthite Under External Magnetic Field

Myung-Hwan Whangbo[a,b,*], Hyun-Joo Koo[b], Eva Brücher[c], Pascal Puphal[c], and Reinhard K. Kremer[c,*]

[a] Department of Chemistry, North Carolina State University, Raleigh, NC 27695-8204, USA
[b] Department of Chemistry and Research Institute for Basic Sciences, Kyung Hee University, Seoul 02447, Republic of Korea
[c] Max Planck Institute for Solid State Research, Heisenbergstrasse 1, D-70569 Stuttgart, Germany

**Abstract**

We determined the spin exchanges between the $Cu^{2+}$ ions in the kagomé layers of volborthite, $Cu_3V_2O_7(OH)_2·2H_2O$, by performing the energy-mapping analysis based on DFT+U calculations, to find that the kagomé layers of $Cu^{2+}$ ions are hardly spin-frustrated, and the magnetic properties of volborthite below ~75 K should be described by very weakly interacting antiferromagnetic uniform chains made up of effective S=1/2 pseudospin units. This conclusion was verified by synthesizing single crystals of not only $Cu_3V_2O_7(OH)_2·2H_2O$ but also its deuterated analogue $Cu_3V_2O_7(OD)_2·2D_2O$ and then by investigating their magnetic susceptibilities and specific heats. Each kagomé layer consists of intertwined two-leg spin ladders with rungs of linear spin trimers. With the latter acting as S=1/2 pseudospin units, each two-leg spin ladder behaves as a chain of S=1/2 pseudospins. Adjacent two-leg spin ladders in each kagomé layer interact very weakly, so it is required that all nearest-neighbor spin exchange paths of every two-leg spin ladder remain antiferromagnetically coupled in all spin ladder arrangements of a kagomé layer. This constraint imposes three sets of entropy spectra with which each kagomé layer can exchange energy with the surrounding on lowering the temperature below ~1.5 K and on raising the external magnetic field $B$. We discovered that the specific heat anomalies of volborthite observed below ~1.5 K at $B = 0$ are suppressed by raising the magnetic field $B$ to ~4.2 T, that a new specific heat anomaly occurs when $B$ is increased above ~5.5 T, and that the imposed three sets of entropy spectra are responsible for the field-dependence of the specific heat anomalies.

Keywords: volborthite; deuterated volborthite; kagomé layer; spin exchange; DFT calculations; two-leg spin ladder; spin 1/2 antiferromagnetic uniform Heisenberg chain; entropy driven magnetic phase transitions; magnetic heat capacity; magnetic field-dependence of specific heat

# 1. Introduction

Properties of a magnetic material are described by a spin Hamiltonian defined in terms of a few spin exchange paths between magnetic ions. The repeat pattern of strong spin exchange paths forms a spin lattice, with which the spin Hamiltonian generates the energy spectrum to be utitilized in describing the magnetic properties. Consequently, use of a correct spin lattice is paramount because the nature of the energy spectrum generated by a spin Hamiltonian depends on the nature of the spin lattice chosen. Antiferromagnets possessing a kagomé arrangement of transition-metal magnetic ions M are often believed to be spin-frustrated, hence suppressing a long-range magnetic order, so they are prime candidates that can give rise to exotic magnetic ground states.[1,2] However, spin frustration is not guaranteed per se by, for example, a trigonal, a kagomé or a pyrochlore structural arrangement of magnetic transition-metal ions M because they interact strongly with the surrounding main-group ligands L to form $ML_n$ polyhedra thereby splitting its d-states. What is crucial for geometrical spin frustration is not the geometrical pattern imposed by the arrangement of magnetic ions M, but rather that of the spin exchanges these ions generate with their neighboring ions. The direction-dependence of spin exchanges between magnetic ions M is determined by their magnetic orbitals, namely, the singly-occupied d-states of their $ML_n$ polyhedra.[3,4] The d-states have the d-orbitals of M combined out of phase with the p-orbitals of the ligands L, so they are highly anisotropic in shape. Consequently, even if the magnetic ions form a kagomé arrangement, the spin exchanges between magnetic ions do not necessarily generate the same pattern. The anisotropy of spin exchange is most strongly manifested for a magnetic ion possessing only one magnetic orbital (e.g., S=1/2 ions $Cu^{2+}$ and $Ti^{3+}$), and least strongly for a magnetic ion with five magnetic orbitals (e.g., S=5/2 ions $Mn^{2+}$ and $Fe^{3+}$).

Volborthite, $Cu_3V_2O_7(OH)_2 \cdot 2H_2O$, consisting of $Cu^{2+}$ ($d^9$, S = 1/2) ions in kagomé arrangement, has received much attention over the past decade,[5-11] largely because it is believed to be spin-frustrated.[5,11] The kagomé layers of $Cu^{2+}$ ions are slightly distorted from an ideal kagomé shape with two different crystallographic sites for $Cu^{2+}$ in a ratio 2:1. Below room temperature, each $Cu^{2+}$ ion forms an axially-elongated $CuO_6$ octahedron so that, with the local z-axis taken along the elongated Cu-O bonds, the magnetic orbital of each $CuO_6$ octahedron is the $x^2-y^2$ state contained in the $CuO_4$ equatorial plane, which is quite anisotropic in shape. In volborthite, the kagomé layers of composition $Cu_3O_6(OH)_2$ parallel to the *ab*-plane are pillared by pyrovanadates $V_2O_7$. The voids between the layers provide space for crystal water molecules. Below room temperature, volborthite undergoes two structural phase transitions, one at ~292 K from a *C*2/*c* phase to a *I*2/*a* phase, and the other at about ~155 K from the *I*2/*a* phase to a *P*2$_1$/*c* phase.[6] The latter structural phase transition generates two slightly different kagomé layers, which are slightly different in structure. Below 1.5 K volborthite exhibits magnetic order indicated by two anomalies in the magnetic specific heat.[10]

On the basis of their DFT+U calculations Janson *et al.*[11] described the magnetic properties of volborthite using a trigonal spin lattice made up of pseudo S=1/2 units, i.e., linear spin trimers in which adjacent $Cu^{2+}$ ions are strongly coupled antiferromagnetically. However, with this spin lattice model, it is difficult to explain why volborthite undergoes a magnetic ordering below 1.5 K.[10] Furthermore, the magnetic susceptibility Janson *et al.* calculated for volborthite by their exact diagonalization method for the spin Hamiltonian of the trigonal lattice show a spin gap below the susceptibility maximum. The latter is in sharp contrast to the experimental observation,[5] which reveals that the magnetic susceptibility remains rather high as the temperature is decreased toward 0 K,[5] and that the magnetic entropy removed by the long-range order is small compared to R*ln*2 as is typically found for low-dimensional short-range ordered (SRO) magnetic systems, e.g., for S=1/2 antiferromagnetic uniform Heisenberg (AUH) chains. Then, one might speculate if the magnetic ordering of volborthite observed below 1.5 K is associated with a magnetic ordering of such S=1/2 AUH chains, although it has been believed that volborthite is spin-frustrated.[5,11]

To explore the magnetic order below 1.5 K, especially, its dependence on magnetic field as well as the origin of the apparent finite susceptibility as $T \rightarrow 0$ K, we re-analyze the spin lattice of volborthite by performing an energy-mapping analysis based on DFT+U calculations,[3,4] to find that the magnetic properties of volborthite below ~75 K should be described by two-leg spin ladders with rungs of S=1/2 pseudospin units rather than by a trigonal lattice of S=1/2 pseudospin units as proposed by Janson *et al*.[11] We verify this conclusion by acquiring new magnetic susceptibility data and re-analyzing reported magnetization data of volborthite, to show that the kagomé layer of $Cu^{2+}$ ions is hardly spin-frustrated, and the low-temperature magnetic properties of volborthite should be described by an AUH chain composed of S=1/2 pseudospin units. We characterize the magnetic phase transitions below 1.5 K by carrying out specific heat measurements for $Cu_3V_2O_7(OH)_2 \cdot 2H_2O$ and its deuterated analogue, $Cu_3V_2O_7(OD)_2 \cdot 2D_2O$, under magnetic field $B = 0 - 9$ T. Our work shows that the magnetic ordering of volborthite below 1.5 K is suppressed by field $B > \sim 4.5$ T while a new magnetic ordering takes place when $B > \sim 5.5$ T, and that these field-dependent behaviors of the magnetic ordering originate from the three sets of magnetic entropy spectra of each kagomé layer of $Cu^{2+}$ ions created by topologically-constrained interactions between adjacent two-leg spin ladders.

## 2. Experimental and calculations

To determine the spin exchanges of the *I*2/*a* and *P*21/*c* phases of volborthite, we carried out DFT+U calculations employing the Vienna ab Initio Simulation Package (VASP)[12,13] using the projector augmented wave (PAW)[14,15] method and the PBE[16] exchange-correlation functional.

The electron correlation associated with the 3d states of Cu was taken into consideration by DFT+U calculations, i.e., by performing DFT calculation with an effective on-site repulsion $U_{eff}$ = U – J = 4 and 5 eV added on the magnetic ions $Cu^{2+}$.[17]

Single crystals of volborthite were grown using hydrothermal techniques as described in the literature.[6] Deuterated samples of volborthite were prepared by replacing light by heavy water (isotope enrichment 99.5 %). The magnetic susceptibilities were measured at constant field as a function of the temperature in a Magnetic Property Measurement System SQUID magnetometer (MPMS-XL7, Quantum Design, San Diego, U.S.A.). The specific heats of a collection of oriented crystals were determined using the relaxation method of a 3-He Physical Property Measurement System (PPMS, Quantum Design, San Diego, U.S.A). In order to construct a lattice reference needed to evaluate the magnetic contribution to the specific heat of volborthite, we prepared a polycrystalline sample of the diamagnetic mineral martyite with composition $Zn_3V_2O_7(OH)_2 \cdot 2H_2O$ from an aqueous solution of $NH_4VO_3$ and zinc acetate, $Zn(CH_3CO_2)_2$, following a recipe reported in the literature.[18]

## 3. Results

### 3.1. Spin exchanges and spin lattice of volborthite

We determine the spin exchanges $J_1 – J_5$ defined in **Figure 1a** by performing the energy-mapping analyses[3,4] based on DFT+U calculations (for details of calculations, see the Supplemental Material). Below room temperature down to ~155 K, volborthite has a crystal structure with two equivalent kagomé layers in the unit cell (space group *I2/a*) phase and below 155 K a structure with two slightly nonequivalent kagomé layers of $Cu^{2+}$ ions (space group *P2₁/a*).[5,8] The spin exchanges of the two phases are very similar as summarized in **Table 1**, where only the values obtained with $U_{eff}$ = 4 eV are listed. Those calculated with $U_{eff}$ = 5 eV are given in the Supplemental Material. Janson *et al.*[11] carried out DFT+U calculations using $U_{eff}$ = 8.5, 9.5 and 10.5 eV to determine the spin exchanges of the *P2$_1$/a* structure by fitting the electronic band structure in terms of a tight binding approximation. In the following we first discuss the spin lattice of volborthite suggested by the spin exchanges we obtained. Subsequently, we show that the same picture is obtained from those of Janson *et al.* despite that their spin exchanges are considerably smaller than ours in the strengths of AFM character due to their use of very large $U_{eff}$ values.

As can be seen from **Table 1**, the strongest exchange $J_2$ is AFM and forms isolated linear trimers (**Figure 1a**). The second strongest spin exchange $J_4$ is also AFM but is weaker than $J_2$ by nearly an order of magnitude. All other spin exchanges are weaker than $J_2$ by more than an order of magnitude. Since $J_2$ is much stronger than other spin exchanges, each linear trimer at $T << J_2$ constitutes a pseudospin S=1/2 unit and such units form a triangular lattice, as already pointed out by Janson *et al.*[11] However, we note that the linear trimers interact through the AFM exchange $J_4$ to form two-leg spin ladders with the linear trimers as rungs (**Figure 1b**). This feature is hidden in a kagomé layer because nearest-neighbor spin ladders are geometrically entangled with their superposed legs.

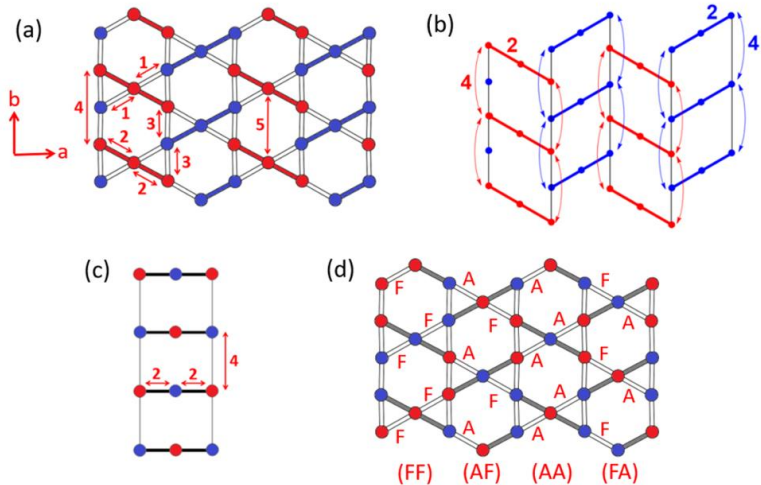

Figure 1. (a) The kagomé arrangement of $Cu^{2+}$ ions (red and blue circles) in volborthite, where the labels 1 – 5 refer to the spin exchanges $J_1$ – $J_5$, respectively (see **Table 1**). In each $J_2$ path, two $Cu^{2+}$ ions make a Cu-O…$V^{5+}$…O-Cu bridge, while the remaining exchange paths do not. (b) Geometrical entanglement of adjacent two-leg spin ladders in a kagomé layer. Adjacent spin ladders are presented with different colors for ease of distinction. (c) A two-leg spin ladder with rungs of linear trimers defined by $J_2$ and legs defined by $J_4$, with all exchange paths atiferromagnetically coupled, where the red and blue circles represent the up-spin and down-spin $Cu^{2+}$ ion sites, respectively. (d) A spin ladder arrangement in which (FF)-, (AF)-, (AA)- and (FA)-coupled spin ladders occur consecutively.

The two-leg spin ladder model as the spin lattice of volborthite described above is also supported by Janson *et al.*'s spin exchanges (**Table 1**), although their spin exchanges differ considerably from ours in magnitude. In general, a spin exchange J between two magnetic ions located at sites i and j can be written as the sum of the FM and AFM components ($J_F$ and $J_{AF}$, respectively), namely, $J = J_F + J_{AF}$.[3,4,19] With the magnetic orbitals describing the spin sites i and j as $\psi_i$ and $\psi_j$, respectively, the overlap density $\rho_{ij} = \psi_i\psi_j$ gives rise to the exchange repulsion $K_{ij}$ while the overlap integral $\langle\psi_i|\psi_j\rangle$ leads to an energy split $\Delta e$ between the two states described by the magnetic orbitals. Then, $J_F$ and $J_{AF}$ are written as

$$J_F = -K_{ij}, \text{ and } J_{AF} = \frac{(\Delta e)^2}{U_{eff}} \qquad (1)$$

Since the AFM component $J_{AF}$ is inversely proportional to $U_{eff}$, using a large $U_{eff}$ value in DFT+U calculations should make *J* value less AFM or even shift it to a FM spin exchange. This explains why the $J_2$ and $J_4$ values of Janson *et al.* are less strongly AFM than ours, and why their $J_1$ and $J_3$ values are more strongly FM than ours.

**Table 1.** Values (in K) of the spin exchanges $J_1$ – $J_5$ calculated for the *I*2/a and *P*2$_1$/a phases of volborthite obtained by DFT+U calculations

|  | *I*2/*a* phase[a] | *P*2$_1$/*a* phase[a] | | *P*2$_1$/*a* phase[b] | |
| --- | --- | --- | --- | --- | --- |
|  |  | Layer 1 | Layer 2 | Layer 1 | Layer 2 |
| $J_1/J_2$ | 0.010 | -0.001 | -0.025 | -0.15 | -0.11 |
| $J_3/J_2$ | 0.025 | 0.031 | 0.031 | -0.34 | -0.32 |
| $J_4/J_2$ | 0.145 | 0.141 | 0.135 | 0.17 | 0.15 |
| $J_5/J_2$ | 0.014 | 0.014 | 0.013 | - | - |
| $J_2$ | 542 K | 550 K | 582 K | 193 K | 205 K |

[a] Present work with $U_{eff}$ = 4 eV

[b] Janson *et al.* (ref. 11) with $U_{eff}$ = 8.5 eV.

We now examine the interactions between adjacent two-leg spin ladders, under the constraint that the spin exchanges $J_2$ and $J_4$ forming the two-leg spin ladders are much stronger than $J_3$ and $J_1$, which provide interactions between spin ladders. The latter requires that, in all possible spin ladder arrangements, all $J_2$ and $J_4$ exchange paths of each two-leg spin ladder must remain antiferromagnetically coupled (**Figure 1d**). In every intertwined legs, each $J_2$ leg has two consecutive $J_3$ paths. Since each $J_2$ leg is antiferromagnetically coupled, $J_3$ has no effect on the interaction between adjacent two-leg spin ladders regardless of whether it is AFM or FM. Thus, adjacent two-leg spin ladders interact only through the very weak exchanges $J_1$ (**Figure 1a**). Note that $J_1$ is weakly AFM for the *I2/a* phase, but is weakly FM for the *P2$_1$/a* phase in our calculations. This difference does not influence how the ordering between spin ladders in a kagomé layer is affected by $J_1$ (see below). Since $J_1$ is more than an order of magnitude weaker than $J_4$, the magnetic property of each kagomé layer at low temperatures ($|J_1| < T \ll J_2$) is primarily determined by those of the two-leg spin ladders with each linear trimer acting as an effective S=1/2 pseudospin rung (**Figure 1c**). At sufficiently low temperatures where excitations within the rungs are negligible, the dominant AFM spin exchanges form two-leg spin ladders predicting that volborthite should be regarded primarily as a system of very weakly coupled S=1/2 AUH chains. In the following we first demonstarte that the magnetic properties of volborthite below ~75 K are well explained by the model of very weakly interacting S=1/2 AUH chains.

## 3.2. Experimental evidence for the spin-half AFM uniform Heisenberg chain character

**Figure 2** displays the magnetic susceptibilities of volborthite, which exhibits a broad peak at ~17 K and a finite susceptibility as $T \to 0$ K characteristic of a low-dimensional AFM SRO behavior. As implied by the results of the DFT+U calculations, the magnetic susceptibilities of volborthite between 3 and 75 K can indeed be very well fitted by those theoretically predicted for a S=1/2 AUH chain according to

$$\chi_{mol}(T) = \chi_{spin}(J_C, g, T) + \chi_0 \qquad (2)$$

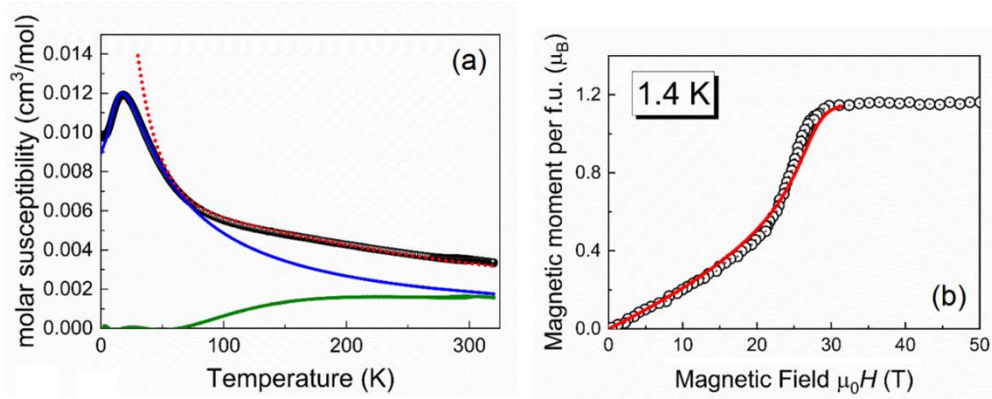

**Figure 2.** (a) The magnetic susceptibility of one formula unit (i.e., comprising three Cu atoms) of volborthite (black circles) with $B$ applied along the crystallographic $b$ axis fitted (for $T < 75$ K) to the theoretical prediction for a S=1/2 AUH chain (solid blue curve) given by eq. (2). The difference between the two is displayed as a solid green line. The experimental susceptibilities for 75 K $\leq T \leq$ 320 K are well fitted by the susceptibility of a linear spin S=1/2 trimer with spin exchange of 197(K) (red dotted curve). Note the susceptibility anomaly at ~295 K due to the structural phase transition from the $C2/c$ to the $I2/a$ phase.[5] (b) Isothermal magnetization of volborthite determined at 1.4 K taken from ref. 11 compared with quantum Monte Carlo calculations for a S=1/2 AUH chain with a spin exchange of $J_C = 27.5$ K (solid red line).

where $\chi_{\text{spin}}(J_C, g, T)$ represents the magnetic susceptibility of the S=1/2 AUH chain with nearest-neighbor (NN) spin exchange $J_C$ (**Figure 2a**). For $\chi_{\text{spin}}(J_C, g, T)$ we use the Padé approximant of Quantum Monte Carlo (QMC) results for the S=1/2 AUH chain.[20] The second term refers to the temperature-independent susceptibility $\chi_0$ arising from the diamagnetism of the electrons in the closed shells (-175 × 10$^{-6}$ cm$^3$/mol per formula unit, FU)[21,22] and also from the van Vleck paramagnetic susceptibility due to excitations to the excited states of each Cu$^{2+}$ ion (~100 × 10$^{-6}$ cm$^3$/mol per Cu atom),[23] leading to $\chi_0 = +125 \times 10^{-6}$ cm$^3$/mol per FU. Up to ~75 K the experimental susceptibility is well reproduced by the susceptibility calculated for a S=1/2 AUH chain with $J_C$ = 27.8(5) K and $g$ = 2.33 (the solid blue curve in **Figure 2a**). The fitted g-factor is found to be 2.33, which is at the higher end of the g-factor expected for Cu$^{2+}$ ions.[24] The S=1/2 AUH chain model readily explains the susceptibility maximum at $T_{\text{max}}$ = 17 K as well as the finite susceptibility below $T_{\text{max}}$. The exchange $J_C$ = 27.8(5) K agrees very well with the expected value $J_C = T_{\text{max}}/0.64085$ = 26 K. In addition, the ratio $\chi_{\text{spin}}T_{\text{max}}/g^2$ = 0.0346 cm$^3$K/mol concurs well with the value 0.0353 expected for the S=1/2 AUH chain.[20] The difference between the experimental data and the chain susceptibility (green solid line in **Figure 1**) vanishes below ~75 K and gradually increases to higher temperatures. The susceptibilities above ~75 K without any constraints matches very well with the susceptibility of an isolated linear spin S=1/2 trimer described by the spin Hamiltonian

$$H = J_{\text{trim}} (\vec{S}_1 \cdot \vec{S}_2 + \vec{S}_2 \cdot \vec{S}_3) \tag{3}$$

where $\vec{S}_i \cdot (i = 1, 2, 3)$ represent the three spin sites. Replacing $\chi_{\text{spin}}(J_C, g, T)$ in eq. (2) with the susceptibility of a spin S=1/2 trimer given, e.g., by Boukhari et al.[25] the susceptibility data above ~75 K can be well fitted without further constraints. The trimer spin exchange amounts to 197(2) K, indicating a ratio $J_C/J_{\text{trim}}$ = 0.14, in good agreement with the ratio $J_4/J_2$ obtained from our DFT+U calculation (see **Table 1**) suggesting that the spin trimers should be identified as the rungs of the two-leg spin ladders running along the crystallographic *b* axis.

Another characteristic feature of the magnetic properties of volborthite is the low-temperature 1/3 magnetization plateau extending over a wide magnetic field range, displayed in **Figure 2b**.[8,9] Note that we multiplied the published literature data by a factor of three to obtain the magnetization per one FU of volborthite (3 Cu atoms), the experimental data are well reproduced by the magnetization predicted for the S=1/2 AUH chain with $J_C = 27.5$ K. The theoretical magnetization was obtained by Quantum Monte Carlo Calculations (QMC) employing the path integral method of the loop code incorporated within the ALPS project.[26,27] To ensure a low statistical error, these simulations included 250 spins and $1.5 \times 10^5$ Monte Carlo steps for the initial as well as a subsequent thermalization.

### 3.3. Specific heat anomalies of volborthite
#### A. $Cu_3V_2O_7(OH)_2 \cdot 2H_2O$

In **Figure 3a**, we show the specific heat measured for an ensemble of aligned crystal of volborthite $Cu_3V_2O_7(OH)_2 \cdot 2H_2O$ in zero magnetic field with magnetic field applied along the *b* axis, the estimated lattice contribution to the heat capacity constructed from the specific heat of $Zn_3V_2O_7(OH)_2 \cdot 2H_2O$,[28] and the difference between the two representing the magnetic contribution to the specific heat of volborthite. It is difficult to obtain a proper lattice specific heat reference because the atomic masses of Cu and Zn are not the same and because the space groups describing the crystal structures of $Zn_3V_2O_7(OH)_2 \cdot 2H_2O$ ($P\bar{3}m1$) and $Cu_3V_2O_7(OH)_2 \cdot 2H_2O$ ($I2/a$ and $P2_1/a$) are different. In our work, we adopt the procedure suggested by Boo and Stout[29] and stretch the temperature axis of the specific heat data for $Zn_3V_2O_7(OH)_2 \cdot 2H_2O$ with a smooth function varying

4from 1.15 at 0 K to 1.05 at 50 K. The latter values were chosen such that the specific heat of $Zn_3V_2O_7(OH)_2·2H_2O$ smoothly approaches that of volborthite as $T \rightarrow 50$ K. The magnetic specific heat of volborthite thus obtained exhibits a broad maximum consistent with the specific heat capacity expected for a S=1/2 AUH with a NN spin exchange of $J_C = 26$ K. The latter value is in good agreement with the spin exchanges derived from the analyses of the magnetic susceptibility and magnetization data. The total magnetic entropy amounts to ~80 % of $R\ln2$ expected for a S=1/2 system, consistent with the result reported by Yamashita et al.[10]

**B. Effect of isotope substitution**

In **Figure 3b**, we compare the specific heats of $Cu_3V_2O_7(OH)_2·2H_2O$ and its deuterated analogue, $Cu_3V_2O_7(OD)_2·2D_2O$, prepared by using 99.5 % isotope enriched heavy water in the preparation. The specific heat anomalies of $Cu_3V_2O_7(OH)_2·2H_2O$ below 1.5 K (see inset **Figure 3b**) are hardly affected by the deuteration. However, the structural phase transition from $C2/c$ to $I2/a$ near 155 K undergoes a 5 % upshift for the deuterated sample. This strongly suggests that the two specific heat anomalies below 1.2 K are related to magnetic ordering. Apparently, the D → H substitution does not essentially modify the spin exchange between adjacent kagomé layers, but substantially affects the structural properties of volborthite. The latter finding reflects the fact that the pyrovanadate groups connecting neighboring kagomé layers provide a large interlayer spacing. The hydrogen bond network, from an OH group of a kagomé layer to a crystal water to an O atom of the adjacent kagomé layer, is loosened by the D → H substitution, which facilitates the $I2/a$ to $P2_1/a$ structural transition in $Cu_3V_2O_7(OD)_2·2D_2O$.

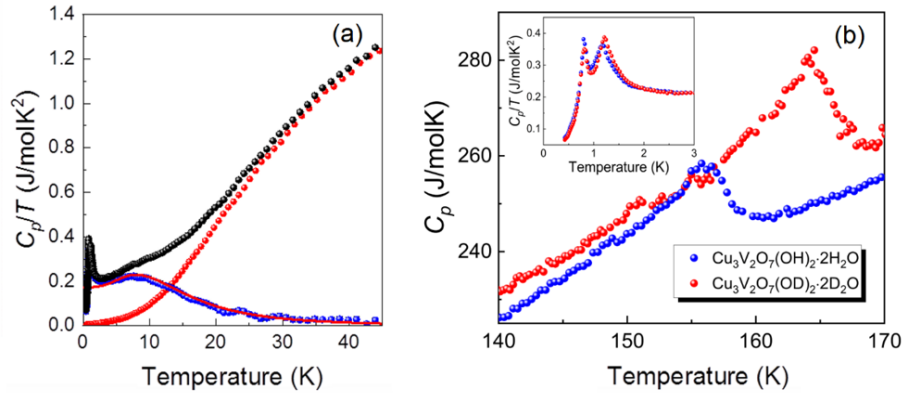

Figure 3. (a) Specific heats of volorthite (black dots) and martyite, $Zn_3V_2O_7(OH)_2·2H_2O$ (red dots). The latter was scaled as described in detail in the text to smoothly approach the heat capacity of volborthite at high temperatures. The difference (blue dots) representing the magnetic contribution to the specific heat of volborthite. The solid red line displays the heat capacity of a S=1/2 AUH with spin exchange of 26 K and spin entropy amounting to ~80 % of $R$ln2 expected for a S=1/2 system. (b) Comparison of the specific heats of $Cu_3V_2O_7(OH)_2·2H_2O$ and $Cu_3V_2O_7(OD)_2·2D_2O$

## C. Effect of magnetic field on magnetic ordering

Given that the specific heat anomalies below 1.5 K are associated with magnetic ordering, it is reasonable to expect that it can be destroyed and the specific heat anomalies disappear under sufficiently high external magnetic field. To confirm this conjecture, we carry out specific heat measurements for volborthite under magnetic fields $B$ = 0 – 9 T (**Figure 4a**). As $B$ increases, the two specific heat anomalies observed at $B$ = 0 widen but remain separated forming two ridges until they eventually merge into one and then abruptly disappear for $B$ > ~4.2 T (see **Figure 4a**). Surprisingly, however, a new specific heat anomaly occurs for fields $B$ > ~5.5 T. This new anomaly forms a single broad ridge, widening and shifting to higher temperatures with increasing $B$.

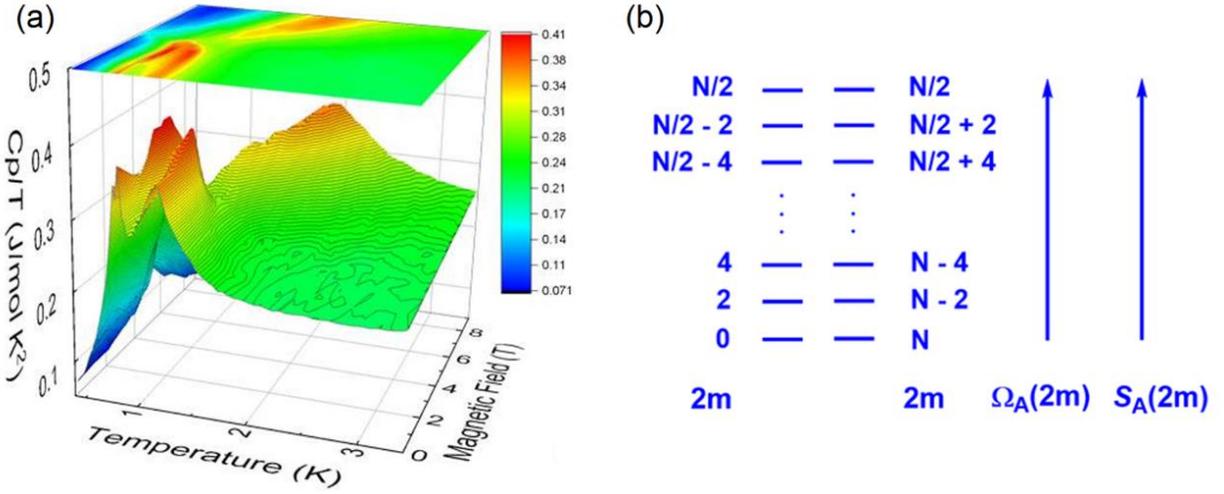

Figure 4. (a) Magnetic field dependence of the specific heat measured with an external magnetic field applied $B$ along the $b$ axis of the volborthite crystals. (b) The spectra of $\Omega_A(2m)$ and $S_A(2m)$ in a kagomé layer of N spin ladders containing m pairs of (AA)- and (FF)-coupled spin ladders. $S_A(2m)$ is the entropy associated with $\Omega_A(2m)$, namely, $S_A(2m) = k_B\ln\Omega_A(2m)$.

To probe the cause for the field-dependence of the specific heat anomalies, it is necessary to find the magnetic energy spectra of volborthite available below 1.5 K with which it can exchange energy with the surrounding phonon bath. At a temperature $T$ below 1.5 K, the free energy $G_i$ of volborthite associated with its magnetic arrangement $i$ can be written as $G_i = H_i - TS_i$, where $H_i$ and $S_i$ are the enthalpy and entropy of the magnetic arrangement $I$, respectively If volborthite has a set of different magnetic arrangements $i = 1, 2, 3, \cdots$, it has the corresponding energy spectrum $\{G_1, G_2, G_3, \cdots\}$. Then, volborthite can absorb energy from the surrounding by using this energy spectrum reflected by the specific heat anomaly. In the next section, we show that volborthite has three distinct but different sets of energy spectra associated with the arrangements of two-leg spin ladders in each kagomé layer with respect to each other The limitation to only three energy spectra occur due to the fact that interactions between adjacent two-leg spin ladders are topologically constrained (see below).

## 4. Entropy spectra and specific heat anomalies

### 4.1. Spin ladder arrangements

In this section we discuss the magnetic-field dependence of the magnetic ordering in volborthite below 1.5 K. Our experiments gave conclusive evidence for an effective AUH chain behavior of volborthite at low temperatures based on the two-leg spin ladders formed by $J_2$ and $J_4$. As already mentioned, the two-leg spin ladders (thereafter, referred to as the spin ladder, for short) interact with their neighboring spin ladders to establish two-dimensional correlations within each kagomé layer via the spin exchange $J_1$ (Section 3.1). To begin with, we examine what kinds of spin ladder arrangements are possible in a kagomé layer. As already mentioned, in all possible magnetic arrangements of each kagomé layer at temperatures below 1.5 K, all $J_2$ and $J_4$ exchange paths of every spin ladder should remain antiferromagnetically coupled (**Figure 1d**) because the spin exchanges $J_2$ and $J_4$ forming the spin ladders are AFM and are much stronger in magnitude than $J_1$. This topological constraint on the magnetic arrangements of a kagomé layer leads to three different groups of spin ladder arrangements. Note that each spin ladder has a set of two consecutive $J_1$ paths running along the leg direction, which consists of the left and right sets of $J_1$ paths. Then, *the first topological constraint* is that the left or right set of $J_1$ paths in each spin ladder be coupled either all antiferromagnetically or all ferromagnetically. For simplicity, we use the labels A (F) to represent the set of $J_1$ paths all antiferromagnetically (ferromagnetically) coupled in each spin ladder. The two sets of the $J_1$ paths in each spin ladder can have one of the four possible arrangements, (**AA**), (**AF**), (**FA**), and (**FF**), as illustrated in **Figure 1d**. Here the (AF) arrangement, for example, means that the spin ladder indicated by a parenthesis has the $J_1$ paths all antiferromagnetically coupled in the left set, but all ferromagnetically coupled in the right set. Use of bold labels in the (**AA**) and (**FF**) arrangements stems from the ease of recognizing the spin ladder arrangements, as will be seen below. **Figure 1d** illustrates a kagomé layer in which the (**FF**), (AF), (**AA**) and (FA) spin ladders occur consecutively. It is convenient to denote this arrangement of spin ladders by (**FF**)(AF)(**AA**)(FA). As can be seen from this arrangement, *the second topological constraint* is that, when crossing from one spin ladder to its adjacent one, the two sets of the $J_1$ paths adjacent to the boundary between the two be opposite in the nature of the spin exchange coupling, e.g., "F)(A" or "A)(F". As a consequence, *the third topological constraint* is that (**AA**) spin ladders cannot be nearest neighbors, nor can be (**FF**) spin ladders. This implies also that (**AA**) and (**FF**) ladders always have to alternate in any spin ladder arrangement. These three constraints imply that there occur only three different sets of spin ladder arrangements in each kagomé layer, in which the members of each set are distinguished by how many pairs of (**AA**) and (**FF**) spin ladders are present. For simplicity, we use $(AF)_m$ to represent the occurrence of $m$ consecutive (AF) spin ladders, and $(FA)_n$ that of $n$ consecutive (AF) spin ladders. Then, with the integers $m$, $n$, $o$, $p$, $q$ and $r$ counting the numbers of consecutive (AF) or (FA) arrangements, examples of the spin ladder arrangements in Groups I, II and III are written as follows:

Group I: $(AF)_m(\mathbf{AA})(FA)_n(\mathbf{FF})(AF)_o(\mathbf{AA})(FA)_p(\mathbf{FF})(AF)_q$.

Group II: $(AF)_m(\mathbf{AA})(FA)_n(\mathbf{FF})(AF)_o(\mathbf{AA})(FA)_p(\mathbf{FF})(AF)_q(\mathbf{AA})(FA)_r$.

Group III: $(FA)_m(\mathbf{FF})(AF)_n(\mathbf{AA})(FA)_o(\mathbf{FF})(AF)_p(\mathbf{AA})(FA)_q(\mathbf{FF})(AF)_r$.

The example of Group I given above illustrates the case when two pairs of (**AA**) and (**FF**) spin ladders are separated by patches of consecutive (FA) and (AF) spin ladders. All possible spin ladder arrangements of Group I are obtained by changing the number of (**AA**) and (**FF**) pairs, m (= 0, 1, 2, 3, ⋯). With respect to any member of Group I, the corresponding member of Group II has one more (**AA**) spin ladder, while that of Group III has one more (**FF**) spin ladder.

**4.2. Entropy spectra**

For a given $(AF)_m(\mathbf{AA})(FA)_n$ arrangement, it is important to note that the $(AF)_m$ patch represents a domain of (AF)-coupled spin ladders while the $(FA)_n$ patch represents a domain of (FA)-coupled spin ladders, and the (**AA**) spin ladder is a boundary between the two domains. Similarly, in a given $(FA)_m(\mathbf{FF})(AF)_n$ arrangement, the (**FF**) spin ladder represents a boundary between the $(FA)_m$ and $(AF)_n$ domains. Thus, each member of Groups I, II and III arrangements is characterized by how many pairs of (**AA**) and (**FF**) boundaries it has, regardless of the widths (e.g., values of $m$, $n$, ⋯) of each individual domain. This point is important in connection with the enthalpy of any given spin ladder arrangement. All members of Group I have an identical enthalpy, $H_I = J_1 \sum_{i>j} \vec{S}_i \cdot \vec{S}_j = 0$ per kagomé layer, because each member has equal numbers of antiferromagnetically- and ferromagnetically-coupled $J_1$ paths. Similarly, all spin ladder arrangements of Group II are identical in enthalpy, $H_{II} = H_I - 2J_1 = -2J_1$ per kagomé layer, and those of Group III are identical in enthalpy, $H_{III} = H_I + 2J_1 = +2J_1$ per kagomé layer. Note that $J_1$ is weakly FM in the $P2_1/a$ phase (see **Table 1**), and that the value of $-2J_1$ for Group II is the enthalpy of one (**AA**) spin ladder while the value of $+2J_1$ for Group III is the enthalpy of one (**FF**) spin ladder. Since $J_1$ is very weak and since a kagomé layer has a large number of spin ladders, one can for all practical purposes assume $H_I = H_{II} = H_{III} = 0$ per kagomé layer, that is, all spin ladder arrangements of Groups I, II and III are practically identical in enthalpy.

The spin ladder arrangements of Groups I, II and III give rise to entropy spectra that are characterized by the number m of (**AA**) and (**FF**) pairs. To probe the entropy spectra of Groups I,

II and III, we consider that a kagomé layer has N spin ladders, where $N = N_A/3$ per FU of volborthite $Cu_3V_2O_7(OH)_2 \cdot 2H_2O$ with $N_A$ as Avogadro's number. The factor 1/3 takes into account the fact that each spin ladder has three $Cu^{2+}$ ions per repeat unit. For the case when there are m pairs of (**AA**) and (**FF**) arrangements, the total number of different spin ladder arrangements, $\Omega_A(2m)$, in Group I, is given by

$$\Omega_I(2m) = 2(_NC_{2m}) = 2(N!)/[(2m)!(N-2m)!],$$

where $_NC_{2m}$ is the binomial expansion coefficient, and the factor 2 arises because the m pairs of sites in each arrangement can be occupied first with a (**FF**) spin ladder or first with an (**AA**) spin ladder when m > 0. For m = 0, $\Omega_I(0) = 2$ because one can have the $(AF)_\infty = \cdots(AF)(AF)(AF)\cdots$ or the $(FA)_\infty = \cdots(FA)(FA)(FA)\cdots$ arrangement. The corresponding numbers for Group II (Group III) are given by

$$\Omega_{II}(2m) = \Omega_{III}(2m) = \Omega_I(2m) \, \Omega_E(2m),$$

where $\Omega_E(2m)$ represents the number of possible sites available for an extra spin ladder after m pairs of (**AA**) and (**FF**) spin ladders are chosen from N spin ladders. Due to the topological constraints on the spin ladder ordering, $\Omega_E(2m) \leq N - 2m$. Then the entropies *S* associated with the three groups of spin ladder arrangements are given by

$$S_I(2m) = k_B \, ln\Omega_I(2m),$$
$$S_{II}(2m) = S_{III}(2m) = S_I(2m) + S_E(2m),$$
$$S_E(2m) = k_B \, ln\Omega_E(2m).$$

Since $\Omega_I(2m) = \Omega_I(N - 2m)$, $\Omega_I(2m)$ increases as 2m increases from 0 toward N/2, reaching a maximum at 2m = N/2, and decreases as 2m increases beyond N/2 toward N (**Figure 4b**). Consequently, $S_I(2m) = S_I(N - 2m)$, and $S_I(2m)$ increases as 2m increases from 0 toward at N/2 and as 2m decreases from N toward N/2.

The kagomé layers of volborthite satisfy the three conditions necessary for a crystalline solid to exhibit a purely entropy-driven phase transitions:[30-32] (1) It has a set of states identical in enthalpy $H$. (2) The members of this set are grouped into subsets $i = 1, 2, 3, \cdots$, with degeneracies $\Omega_1, \Omega_2, \Omega_3, \cdots$, respectively. (3) The degeneracy $\Omega_i$ increases steadily, e.g., $\Omega_1 < \Omega_2 < \Omega_3 < \cdots$, so the associated entropy increases steadily. Thus, one may expect that already moderate magnetic fields inducing Zeeman energies comparable to $J_1$ are able to substantially alter the entropy balance causing dramatic effects on the low-temperature ordering behavior. In fact, as shown **Figure 4a**, the specific heat anomalies observed at $B = 0$ indeed disappears for $B > \sim 4.2$ T, but a new anomaly occurs when $B > \sim 5.5$ T. Below ~2 K, neither electronic nor vibrational energy of volborthite is available for the energy absorption causing the specific heat anomalies. Thus, in the following, we analyze the dependence of the specific heat anomalies on $T$ and $B$ in terms of the entropy spectra of Groups I, II and III (**Figure 4b**).

### 4.3. Specific heat anomalies and entropy spectra

As shown above, the entropy $S_I(2m)$ of Group I is smaller than that of Group II or III by $S_E(2m)$. As the magnetic field $B$ increases from 0, a kagomé layer starts absorbing energy by exploiting the entropy spectrum of Group I, so the anomaly below ~4.2 T arises from the entropy spectrum of Group I. Since $\Omega_I(2m)$ becomes larger with increasing 2m from 0 to N/2 and also with decreasing 2m from N to N/2, more states are involved in the energy absorption from Zeeman energy. This allows one to understand why the intensity of the specific heat anomaly increases with increasing field $B$. The entropies $S_I(2m)$ from the states for 2m = 0 – N/2 as well as 2m = N – N/2 are used to absorb energy, eventually reaching the highest entropy level $S_I(N/2)$ (**Figure 4b**). When all entropy levels are exploited, there is no more entropy level with which a kagomé layer can absorb energy any further even if $B$ is increased. At this stage, the spin ladder ordering within a kagomé layer assumes a liquid-like disordered state, and the specific heat in the valley between ~4.2 T and ~5.5 T reaches that of the disordered state of the S=1/2 AUH chains (the green plateau in **Figure 4a**). It is most reasonable to assign the higher-temperature specific heat ridge to a magnetic ordering within each kagomé layer, and the lower-temperature one to a three-dimensional ordering, i.e., an ordering between the magnetically-ordered kagomé layers. When the magnetic field increases further, a kagomé layer can utilize the entropy spectra of Groups II and III simultaneously, leading to the single broad specific heat anomaly above ~5.5 T. However, this requires that the spin ladder arrangements of a kagomé layer must be converted from Group I to Groups II and III. Each spin ladder has two sets of $J_1$ exchanges, e.g., a (AF) spin ladder has one set of AFM-coupled $J_1$ and another set of FM-coupled $J_1$ exchanges. Suppose that a set of $J_1$ paths is switched from FM to AFM, e.g., from an (AF) to a (**AA**) spin ladder in one member of $\Omega_A(2m)$ of Group I to create an extra (**AA**) spin ladder. Such a $J_1$-flip, e.g., $(AF)_m(AF)(AF)_n \rightarrow (AF)_m(\mathbf{AA})(FA)_n$, converts the electronic structure from Group I to Group II, raising the entropy from $S_I(2m)$ to $S_{II}(2m) = S_I(2m) + S_E(2m)$. Similarly, a $J_1$-flip can create an extra (**FF**) spin ladder, e.g., $(FA)_m(FA)(FA)_n \rightarrow (FA)_m(\mathbf{FF})(AF)_n$, converting the electronic structure from Group I to Group III, raising the entropy from $S_I(2m)$ to $S_{III}(2m) = S_I(2m) + S_E(2m)$.

With magnetic field, it is energetically more favorable to induce a $J_1$-flip from AFM to FM rather than that from FM to AFM. In contrast, thermally driven $J_1$-flips from AFM to FM or from FM to AFM are equally likely. Thus, thermal agitation will be more effective in creating $J_1$-flips than Zeeman energy, approximately by a factor of 2. The positions of two ridges of the specific heat below ~4.2 T are nearly parallel to the magnetic field axis, whereas the single specific heat ridge above ~5.5 T is slanted with slope $g\mu_B B/k_B T \approx 2$. This reflects that, at strong enough magnetic field, the kagomé layers of Group II spin ladder arrangements absorb energy equally as do those of Group III spin ladder arrangements. The entropy spectra II and III are identical in the degeneracy of each entropy level, but differ in the nature of spin ladder arrangements. Whether a given kagomé layer produces the entropy spectrum II or III has no effect on the enthalpy difference. Consequently, there is no ordering between kagomé layers, which explains the occurrence of a single specific heat anomaly at magnetic field above −5.5 T. The latter forms a single ridge with increasing B, the shape of which suggests that not all available entropy states of the entropy spectra II and III are populated at 9 T. If it is assumed that this above-5.5T specific heat anomaly (**Figure 4a**) retains a symmetric shape with respect to temperature and magnetic field, one might expect that the magnetic order above ~5.5 T will disappear under magnetic field higher than ~13 T.

Finally, we note that in all ordered magnetic states of Groups I, II and III, there occurs a strong gain of translational entropy with respect to those states in which there is no order between neighboring spin ladders, because the ordered states have translational symmetry along the leg direction. This provides a strong driving force for the magnetic ordering. In a sense, this situation is similar to the nematic phase transition in a system of thin rods examined 6 by Onsager.[31]

## 5. Concluding remarks

The spin exchanges of volborthite show that each kagomé layer of $Cu^{2+}$ ions is hardly spin-frustrated, but rather consists of very weakly interacting two-leg spin ladders with linear trimers as rungs. Below ~75 K, these rungs act as S=1/2 pseudospin units, making each two-leg spin ladder behave as a S=1/2 antiferromagnetic uniform Heisenberg chain. This conclusion was confirmed by synthesizing single crystal samples of volborthite $Cu_3V_2O_7(OH)_2 \cdot 2H_2O$ and its deuterated analogue $Cu_3V_2O_7(OD)_2 \cdot 2D_2O$ and subsequently measuring their magnetic susceptibilities and specific heat anomalies. Under magnetic field $B$ higher than ~4.2 T, the specific heat anomalies below 1.5 K are suppressed. However, a new specific heat anomaly appears when $B$ is raised above ~5.5 T. The dependence of the specific heat anomalies on magnetic field are governed by the fact that the magnetic properties of each kagomé has three sets of entropy spectra with which it can exchange energy with the surrounding. These entropy spectra arise from the topologically-constrained interactions between adjacent two-leg spin ladders in each kagomé layer that all spin exchange paths forming each two-leg spin ladder should remain antiferromagnetically coupled.

      The present work makes it clear that use of a correct spin lattice is critical in describing the properties of a magnetic material. As a criterion for finding a proper spin lattice, the geometrical pattern of magnetic ion arrangement can be misleading because a spin lattice is decided by the geometrical pattern of the strong spin exchange paths.[3,4,19] If volborthite were to be treated as a spin-frustrated kagomé lattice model, the S=1/2 antiferromagnetic uniform Heisenberg chain behaviors of its magnetic susceptibility and magnetization are exotic and novel, and so are its specific heat anomalies below 1.5 K. By the same token, all other magnetic properties of volborthite not explained by a spin-frustrated kagomé lattice model would be novel and surprising. Conversely, each kagomé layer of $Cu^{2+}$ ions consists of very-weakly interacting two-leg spin ladders, so the seemingly exotic magnetic properties previously attributed to magnetic frustration are simply explained by a well-studied S=1/2 antiferromagnetic uniform Heisenberg chain model. The most unbiased, straightforward way to find a correct spin lattice for any given magnetic material is to evaluate the relative strengths of various possible spin exchanges of a given magnetic system by using the energy-mapping analysis based on first principles DFT calculations.[3,4,19,33]

**Acknowledgments**

The work at KHU was supported by the Basic Science Research Program through the National Research Foundation of Korea (NRF) funded by the Ministry of Education (2020R1A6A1A03048004).

**Competing interests**

The authors declare no competing interests.

Supplementary Material

for

**Absence of Spin Frustration in the Kagomé Layers of $Cu^{2+}$ Ions in Volborthite $Cu_3V_2O_7(OH)_2 \cdot 2H_2O$ and Observation of the Suppression and Re-entrance of Specific Heat Anomalies in Volborthite Under External Magnetic Field**

Myung-Hwan Whangbo[a,b,*], Hyun-Joo Koo[b], Eva Brücher[c], Pascal Puphal[c], and Reinhard K. Kremer[c,*]

[a] Department of Chemistry, North Carolina State University, Raleigh, NC 27695-8204, USA

[b] Department of Chemistry and Research Institute for Basic Sciences, Kyung Hee University, Seoul 02447, Republic of Korea

[c] Max Planck Institute for Solid State Research, Heisenbergstrasse 1, D-70569 Stuttgart, Germany

**Spin exchanges of the *I2/a* and *P2₁/a* phases of volborthite by energy mapping analysis**

To extract the values of the spin exchanges for the *I2/a* and *P2₁/a* phases of volborthite, we carry out DFT calculations encoded in the VASP[1,2] using the augmented plane wave method[3,4] and the PBE exchange-correlation functional.[5] To take into account the effect of electron correlation of the Cu 3d states, we use the DFT plus on site repulsion (DFT+U) method[6] with $U_{eff} = U - J = 4$ and 5 eV.

**A. *I2/a* phase**

To determine the five spin exchanges $J_1 - J_5$ of the *I2/a* phase, we consider six ordered spin states FM, AF(i) (i = 1 to 5) shown in **Figure S1**. Then, the total spin exchange energies of these states can be written as

$$E = \sum_{i=1}^{5} n_i J_i S^2 \qquad (S1)$$

where S refers to the spin of the $Cu^{2+}$ ion (i.e., S = 1/2). The values of $n_i$ ( i = 1 to 5) found for the six spin states are listed in **Table S1**. The relative energies (meV/FU) obtained for the FM, and AFi (i = 1 – 5) states by DFT+U calculations are listed in **Table S2**. By mapping the relative energies of the ordered magnetic states determined by DFT+U calculations to those determined by the spin exchange energies, we obtain the values of the spin exchanges $J_1 - J_5$ listed in **Table S3** and **Table 1** of the text.

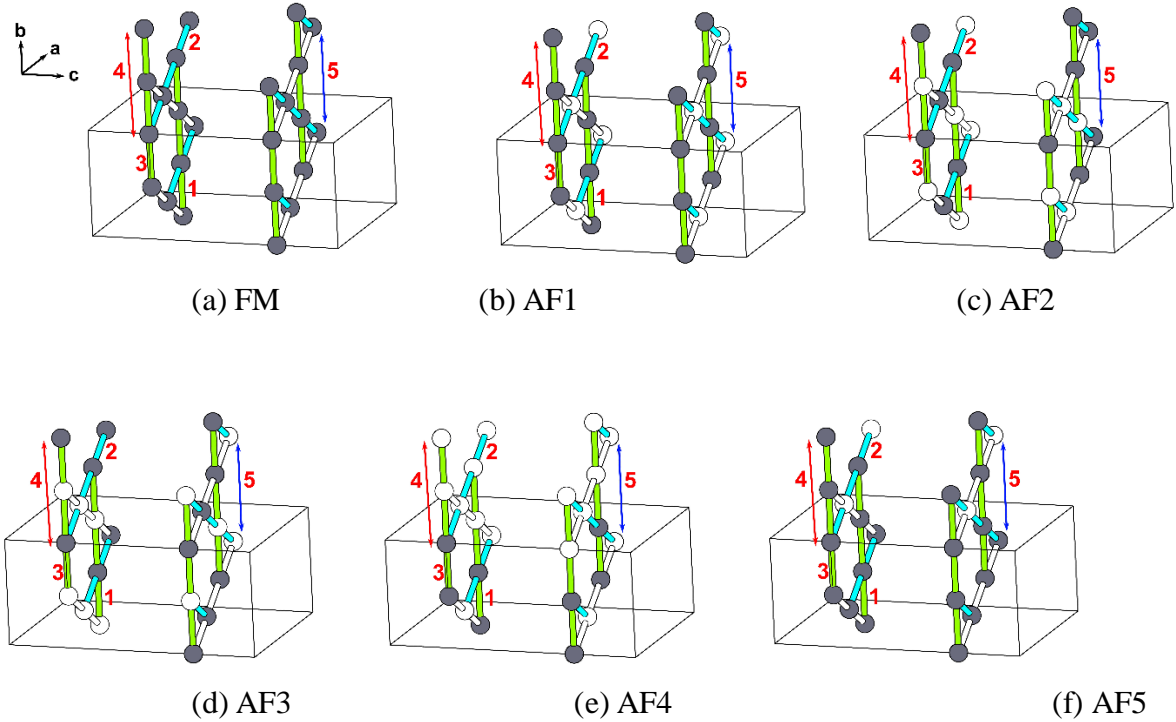

Figure S1. Six ordered spin states, i.e., the FM and AF(i) (i = 1 to 5) states, of the $I2/a$ phase employed for the energy-mapping analysis, where the gray and white circles represent the up and down spin sites of $Cu^{2+}$ ions.

Table S1. Values of $n_i$ for the six ordered spin states

|  | $n_1$ | $n_2$ | $n_3$ | $n_4$ | $n_5$ |
|---|---|---|---|---|---|
| $E_{FM}$ | -16 | -16 | -16 | -16 | -8 |
| $E_{AF1}$ | 16 | 16 | -16 | -16 | -8 |
| $E_{AF2}$ | -16 | 16 | 16 | -16 | -8 |
| $E_{AF3}$ | 16 | -16 | 16 | -16 | -8 |
| $E_{AF4}$ | 0 | 0 | 0 | 16 | -8 |
| $E_{AF5}$ | 0 | 0 | -16 | -16 | 8 |

Table S2. Relative energies (meV/FU) obtained from DFT+U calculations

|  | $U_{eff}$ = 4 eV | $U_{eff}$ = 5 eV |
|---|---|---|
| FM | 47.87 | 37.85 |
| AF1 | 0.69 | 0 |
| AF2 | 0 | 0.65 |
| AF3 | 46.24 | 38.49 |
| AF4 | 16.94 | 13.61 |
| AF5 | 23.95 | 18.63 |

Table S3. Values of the spin exchanges $J_1 - J_5$ (in K) calculated for the $I2/a$ phase

|  | $U_{eff}$ = 4 eV | $U_{eff}$ = 5 eV |
|---|---|---|
| $J_1/J_2$ | 0.010 | 0 |
| $J_3/J_2$ | 0.025 | -0.017 |
| $J_4/J_2$ | 0.145 | 0.149 |
| $J_5/J_2$ | 0.014 | 0.016 |
| $J_2$ | 542 K | 439 K |

**B. $P2_1/a$ phase**

To determine the 10 spin exchanges of the $P2_1/a$ phase, namely, $J_1 - J_5$ for the layer 1 and $J'_1 - J'_5$ for the layer 2, we consider 11 ordered spin states FM, AF(i) (i = 1 to 10) shown in **Figure S2**. Then, the total spin exchange energies of these states can be written as

$$E = \sum_{i=1}^{5} n_i J_i S^2 + \sum_{i=1}^{5} n'_i J'_i S^2 \tag{S2}$$

where $S = 1/2$. The values of $n_i$ (i = 1 to 5) and $n'_i$ (i = 1 to 5) found for the 11 spin states are listed in **Table S4**. The relative energies (meV/FU) obtained for the FM, and AFi (i = 1 – 10) states by DFT+U calculations are listed in **Table S5**. By mapping the relative energies of the ordered magnetic states determined by DFT+U calculations to those determined by the spin exchange energies, we obtain the values of the spin exchanges listed in **Table S6** and **Table 1** of the text.

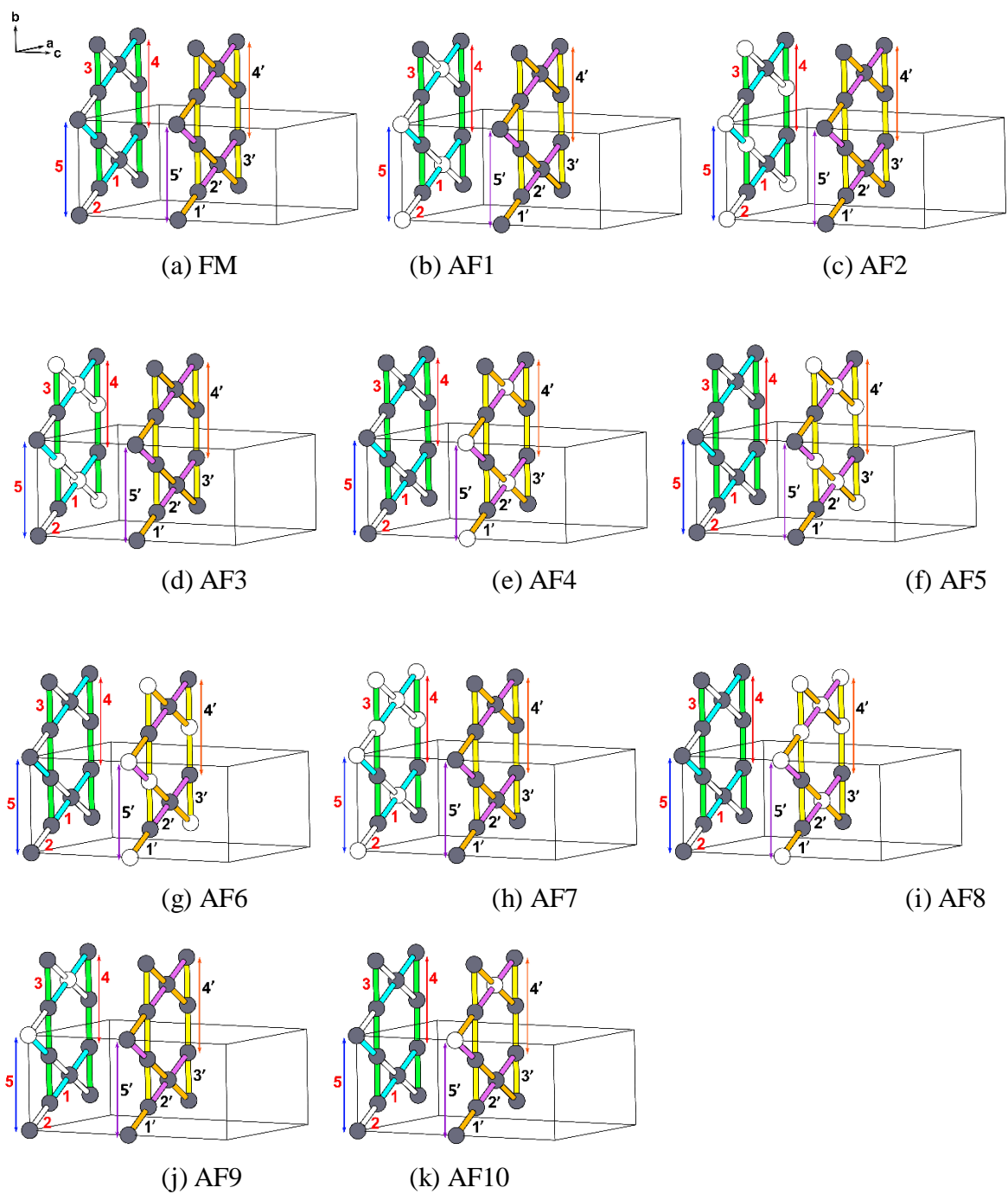

Figure S2. 11 ordered spin states, i.e., the FM and AF(i) (i = 1 to 10) states, of the $P2_1/a$ phase employed for the energy-mapping analysis, where the gray and white circles represent the up and down spin sites of $Cu^{2+}$ ions.

Table S4. Values of $n_i$ and $n_i'$ for the 11 ordered spin states

|  | $n_1$ | $n_2$ | $n_3$ | $n_4$ | $n_5$ | $n_1'$ | $n_2'$ | $n_3'$ | $n_4'$ | $n_5'$ |
|---|---|---|---|---|---|---|---|---|---|---|
| $E_{FM}$ | -8 | -8 | -8 | -8 | -4 | -8 | -8 | -8 | -8 | -4 |
| $E_{AF1}$ | 8 | 8 | -8 | -8 | -4 | -8 | -8 | -8 | -8 | -4 |
| $E_{AF2}$ | -8 | 8 | 8 | -8 | -4 | -8 | -8 | -8 | -8 | -4 |
| $E_{AF3}$ | 8 | -8 | 8 | -8 | -4 | -8 | -8 | -8 | -8 | -4 |
| $E_{AF4}$ | -8 | -8 | -8 | -8 | -4 | 8 | 8 | -8 | -8 | -4 |
| $E_{AF5}$ | -8 | -8 | -8 | -8 | -4 | -8 | 8 | 8 | -8 | -4 |
| $E_{AF6}$ | -8 | -8 | -8 | -8 | -4 | 8 | -8 | 8 | -8 | -4 |
| $E_{AF7}$ | 0 | 0 | 0 | 8 | -4 | -8 | -8 | -8 | -8 | -4 |
| $E_{AF8}$ | -8 | -8 | -8 | -8 | -4 | 0 | 0 | 0 | 8 | -4 |
| $E_{AF9}$ | 0 | 0 | -8 | -8 | 4 | -8 | -8 | -8 | -8 | -4 |
| $E_{AF10}$ | -8 | -8 | -8 | -8 | -4 | 0 | 0 | -8 | -8 | 4 |

Table S5. Relative energies (meV/FU) obtained from DFT+U calculations

|  | $U_{eff} = 4$ eV | $U_{eff} = 5$ eV |
|---|---|---|
| FM | 25.87 | 20.25 |
| AF1 | 2.18 | 1.21 |
| AF2 | 1.42 | 1.21 |
| AF3 | 25.14 | 20.61 |
| AF4 | 1.41 | 0.58 |
| AF5 | 0 | 0 |
| AF6 | 25.72 | 21.08 |
| AF7 | 10.31 | 8.03 |
| AF8 | 9.86 | 7.65 |
| AF9 | 13.86 | 10.59 |
| AF10 | 13.49 | 10.27 |

35Table S6. Values of the spin exchanges $J_1 - J_5$ (in K) calculated for the $P2_1/a$ phase

|  | $U_{eff} = 4$ eV | | $U_{eff} = 5$ eV | |
| --- | --- | --- | --- | --- |
|  | Layer 1 | Layer 2 | Layer 1 | Layer 2 |
| $J_1/J_2$ | -0.001 | -0.025 | -0.009 | -0.035 |
| $J_3/J_2$ | 0.031 | 0.031 | -0.009 | -0.006 |
| $J_4/J_2$ | 0.141 | 0.135 | 0.145 | 0.138 |
| $J_5/J_2$ | 0.014 | 0.013 | 0.015 | 0.014 |
| $J_2$ | 550 K | 582 K | 446 K | 473 K |